\documentstyle[aps,prl,graphicx,floats]{revtex}

\begin{document}
\draft
\twocolumn[\hsize\textwidth\columnwidth\hsize\csname
@twocolumnfalse\endcsname

\title{Effects of pressure on diffusion and vacancy formation in MgO
from non-empirical free-energy integrations.}
\author{Joel Ita and Ronald E. Cohen}
\address{Geophysical Laboratory and Center for High Pressure\\
Research, Carnegie Institution of Washington, 5251 Broad Branch Road,\\
NW, Washington, DC 20015-1305}

\date{\today}

\maketitle

\begin{abstract}
The free energies of vacancy pair formation and migration in MgO
were computed via molecular dynamics using free-energy
integrations and a non-empirical ionic model with no adjustable
parameters.  The intrinsic diffusion
constant for MgO was obtained at pressures from 0 to 140
GPa and temperatures from 1000 to 5000 K. Excellent
agreement was found with the zero pressure diffusion data within 
experimental error. The homologous temperature model which relates 
diffusion to the melting curve describes well our high
pressure results within our theoretical framework.
\end{abstract}


\vskip2pc]
\narrowtext

Diffusion and vacancy formation are critical to kinetic
processes in materials, yet little is known about diffusion at
ultra-high pressures due to experimental difficulties.  Rheology of
oxide minerals at high pressures is also crucial in geophysics
and is dependent on diffusive behavior which is only available 
experimentally at relatively low pressures\cite {karato92}. 
In ionic systems such as MgO, the dominant intrinsic defect is 
the pair vacancy\cite {tilley,ando,yang-flynn} with Mg
and O sites vacant.
Previous work on vacancies in MgO used pseudopotential computations
\cite{devita} or
lattice dynamics or the Mott-Littleton approach with a variety of
semi-empirical potentials\cite{vocal-pri,mac-stew} . The accuracy of 
quasiharmonic lattice dynamics calculations 
degrades above the Debye temperature and the Mott-Littleton
procedure and pseudopotential computations were restricted to 0 K. 

We used molecular dynamics (MD) with non-empirical potentials to determine
the self-diffusion coefficient D where\cite{tilley}
\begin{equation}
D=Z_{f}\frac{Z_{m}}{6}l^{2}\nu \exp (\frac{\frac{\Delta G_{f}}{W}+\Delta
G_{m} }{k_{b}T})  \label{diff-rel}
\end{equation}
$Z_{f}$ is the number of equivalent ways of forming a vacancy
type, $Z_{m}$ is the number of equivalent diffusion paths, 
$l$ is the jump distance, $\nu $ is the attempt frequency, $\Delta G_f$ and
$\Delta G_m$ are the energies of formation and migration, respectively and
$W$ is the solubility factor for polyatomic materials. If the sites are
uncorrelated (Schottky defect), then,
for rocksalt structured (B1) crystals such as MgO, $W=2$, $Z_{f}$ = 1, 
$Z_{m}$ = 12. Highly correlated defects (bound pair) require 
$W$ =1 and $Z_{f}$ = 6.  
Symmetry and energy considerations determine
the value of $Z_{m}$. In either
case, $l^{2}$ = $a^2/2$ where $a$ is the cubic cell parameter.

We used the variational induced breathing (VIB) model which 
reliably gives the thermal properties and 
equation of state of $MgO$\cite{inbar} to compute the energetics 
and interatomic forces. The VIB model is a
Gordon-Kim type model\cite{gordon} in which the total charge density
is modeled by overlapping ionic charge densities which are computed
using the local density approximation (LDA)
\cite{cohen-boy}. The total energy is a sum of three
terms: (a) the long-range electrostatic energy computed using the Ewald 
method, (b) the self-energy
of each atom and (c) the short range interaction energy, the
sum of the kinetic, short-range electrostatic and
exchange-correlation energies from the LDA. There are three
approximations beyond the LDA\cite{hedin}:
(1) The charge density is modeled rather than
computed self-consistently. Comparisons with accurate linearized
augmented plane wave (LAPW) computations show this is a good
approximation for MgO\cite{mehl}; (2) the pair approximation is
used for the short-range interactions (c) which is a good
approximation as long as closed shell ions are used\cite{boyer};
(3) the Thomas-Fermi kinetic energy is used for the short-range
overlap kinetic energy.  The self-energy (b) includes the correct
LDA Kohn-Sham kinetic energy. O$^{2-}$ is not stable in the free
state and is stabilized by introducing a sphere of $2+$
charge (Watson sphere) around it in the LDA atomic
calculations. Interactions are obtained for overlapping ion pairs at
different distances with different Watson sphere radii on the O's.
For efficiency, the interactions were fit with a 21 parameter
analytical expression as functions of $r$, the interatomic distance, 
and $U_i=z_i/R_i$ where $U_i, z_i$, and $R_i$ are the Watson 
sphere potential, charge (2+) and radius for atom i, respectively.
During the simulations, the total
energy was variationally optimized with respect to all of the Watson
sphere radii at each time step.  

The attempt frequency $\nu$ was determined by
Fourier transforming the trajectories of the diffusing ion
projected onto the shortest path to the vacancy. We considered
two models: (1) the lowest frequency
peak in the spectrum, assuming that the diffusive motion is mostly from
the lowest energy mode and (2) the average frequency computed
from the Fourier transform.  In the first case we found that $\nu =$ 
5.2 THz and is independent of $P$ and $T$ over the range studied.
The second case gave an attempt frequency that is within a factor 2 
of the low frequency value . Given the uncertainties
in the calculations and experimental determinations, the difference in the
final results between these two approaches is small and we adopted
case (1) below.

Free energies were computed with the finite time variational or ``adiabatic
switching'' thermodynamic integration method\cite{hunter-rein}. The free
energy difference between the initial and final state is
\begin{equation}
\Delta F=\int_{0}^{1}\frac{\partial F(\lambda )}{\partial \lambda }d\lambda
~=~\int_{0}^{1}\langle \frac{\partial H(\lambda )}{\partial \lambda }\rangle
_{\lambda }d\lambda  \label{del-helm}
\end{equation}
where $\lambda $ is a progress variable which ranges from 0 to 1
as the system ``switches'' from its initial to final state, $H$ is the
system Hamiltonian, and $\left\langle {}\right\rangle _{\lambda }$ 
represents an ensemble average.
In order to obtain $\Delta G_f$, we first calculated 
the free energy difference between an ideal crystal at volume, $V_I$, 
giving the desired average $P$ at $T$ and 
an Einstein crystal at the same $V$ and $T$. This was repeated
for a defective crystal with a bound vacancy pair in each periodic cell 
at $V_D$ corresponding to $P$.
Then for an $N$ atom periodic cell
\begin{equation}
\begin{array}{c}
\Delta G_{f}=F_{D}^{N-L} (V_{D}) - \frac{N-L}{N} F_{I}^{N} (V_{I}) \\ 
+ P \left[ V_{D} - \frac{N-L}{N} V_{I} \right] 
\end{array}
\label{form-free}
\end{equation}
where $F_{D}^{N-L}(V_{D})$ is the Helmholtz free energy of a
defective crystal with $L$ vacant sites and $F_{I}^{N}(V_{I})$ 
is the Helmholtz free energy of the ideal crystal. The
Hamiltonian took the form 
\begin{equation}
H(\lambda )=H_{\text{VIB}}\times (1-\lambda )+H_{ein}\lambda \text{.}
\label{ham-swi}
\end{equation}
where $H_{ein}$ is the Hamiltonian for an Einstein crystal\cite{koning}
which can be written as 
\begin{equation}
H_{ein}=K+U_{o}+\sum_{i=1}^{N}\frac{1}{2}m_{i} \omega_{ein,i}(%
\overrightarrow{x_{i}}-\overrightarrow{x_{i0}})^2  \label{Ein-ham}
\end{equation}
where $K$ is the kinetic energy, $U_{o}$ is the static contribution to the
potential, $m_{i}$, $\overrightarrow{x_{i}}$, $\overrightarrow{x_{i0}}$, and 
$\omega _{ein,i}$ are the mass, position, static lattice position, and
Einstein frequency of the $i$th particle, respectively. The form of 
$\lambda $ as a function of the scaled time, $\tau $, is 
\begin{equation}
\lambda (\tau )=\tau ^{5}(70\tau ^{4}-315\tau ^{3}+540\tau ^{2}-420\tau +126)
\label{swi-func}
\end{equation}
where $\tau =t/t_{s}$, $t$ is the elapsed time, and $t_{s}$ is the total
switching time\cite{koning}.

Migration free energies were calculated using the adiabatic switching
procedure at constant $P$ and $T$. We computed the energy it takes 
to push the atom out of one lattice site and into another vacant 
lattice site\cite{heine}. The force on the migrating atom due to $ H_{VIB} $
in the migration direction was set to zero and the negative of
this force was evenly distributed among the rest of the atoms so that the
force on the center of mass was zero. The position of
the migrating atom was then incremented in the migration direction. 
Forces on the the atom in
the plane perpendicular to the migration direction were not artificially
constrained so that the path the migrating atom took did not
lie on a direct line between the initial position of the atom and the
vacancy site. $\Delta G_{m}$ was determined by summing
the difference in $H_{\text{VIB}}$ before and after incrementing the
position of the migrating atom with the other atomic positions held fixed.
We found that the barriers to migration for ions that do not have a
vacancy as a nearest neighbor were lower than for ones who
do. Thus the value of $Z_{m}$ is 8 in MgO.

MD was performed using a timestep of 1
fs with a 5th order Gear predictor-corrector
scheme\cite{gear} in an isobaric-isothermal ensemble generated using
the extended system method\cite {martyna} for 10
ps equilibration times followed by a 10(15) ps switching time for the
formation(migration) energy. Convergence with respect to our nominal
216 atom system size was verified for systems with up to 1000 atoms 
to be within 1\%. Doubling the integration time resulted in free energy
variations of 1\% while halving it increased the calculated
free energy change by 10\%. The computationally
efficient first principles method used here lends itself to
demanding convergence tests, especially with respect to system size,
that would prove too time-consuming with self-consistent methods\cite{devita}.

\begin{figure}
\includegraphics[clip]{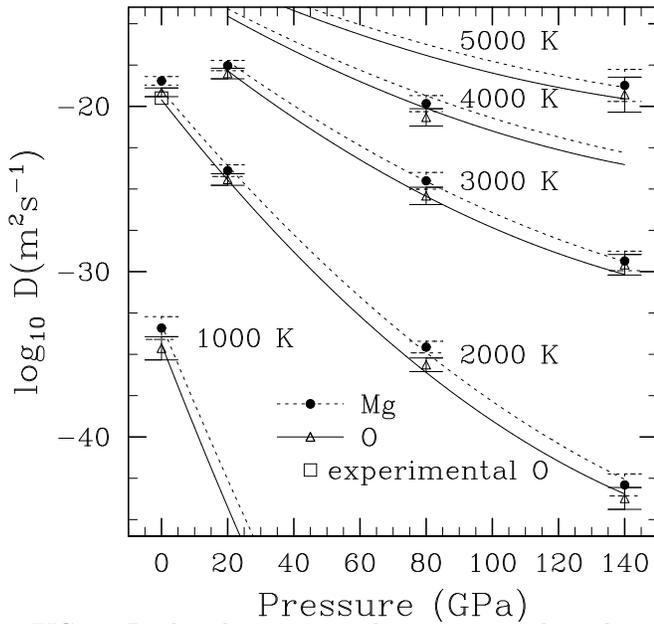}
\caption{Predicted pressure and temperature dependence of the 
self-diffusion coefficients in MgO. Curves represent the best fit to
the coefficients using the activation energy-volume relation given
by Equation \ref{avol}. Vertical symbol size of experimental datum 
taken from Ref. \protect\onlinecite{yang-flynn} represents
uncertainty. \label{fig-diff}}
\end{figure}

Values of $\Delta G_f$ for bound pairs and $\Delta G_m$ are given in 
Table \ref{fe-dif}.
At 0 GPa, these energies are within
5\% of those derived from previous theoretical and experimental results
\cite{ando,yang-flynn,vocal-pri}.
To determine the dominant vacancy mechanism, we calculated the binding 
energy $\Delta G_b$ of a bound pair from the difference in 
$\Delta G_f$ between bound vacancy pairs and those with the largest
possible distance between vacant Mg and O sites in a 1000 atom supercell
corrected for image forces\cite{mills} and found $\Delta G_b$ = 
2.5$\times 10^{-19}$J at 0 GPa and 2000 K and 
7.0$\times 10^{-19}$J at 140 GPa and 3000 K.
No significant changes in migration
energy relative to the bound pair simulation were found. 
Assuming that the variation in binding energy is
linear in pressure and independent of temperature and taking into account
the configurational entropy, we calculated the vacancy concentration and
Gibbs free energy change for crystals containing bound and
disassociated pairs relative to the perfect crystal\cite{tilley}.
This analysis shows that Schottky defects dominate at 1000 K or above
due to entropy contributions (lower temperatures will favor the bound state).
Ionic conductivity measurements indicate that Mg diffusion ($D_{Mg}$)
is controlled by impurities \cite{semp-king} whereas O diffusion
($D_O$) is intrinsic in nature and directly comparable to our analysis. We
find that our predicted $D_O$ agrees with experiment within
mutual error (Fig.~\ref{fig-diff}).

Fitting our diffusion constants with the relation 
\begin{equation}
\begin{array}{c}
\ln D ~=~ \ln ( a^{2} \nu ) + S^*_o + P S^{* \prime}_o \\
- (E^*_o + P V^*_o + P^2 V^{* \prime}_o )/k_b T
\end{array}
\label{avol}
\end{equation}
gives the zero pressure activation entropy $S^*_o =$3-(4)$k_b$, its pressure
derivative $S^{* \prime}_o =$0.03-(0.02)$k_b$, activation energy 
$E^*_o =$9.0-(9.4) $\times 10^{-19}$ J, activation volume $V^*_o=$16.0-(16.7) 
\AA$^3$, and its pressure derivative 
$V^{* \prime}_o =$-0.031-(-0.038) \AA$^3$/GPa 
for Mg-(O). The activation volume varies as a function of pressure consistant 
with previous discussions\cite{mills,karato1,poi-lib}. The activation entropy
is of the same order as previous estimates of the formation entropy at 0
GPa but varies much less drastically with pressure than a previous
estimate\cite{mills}.

Finally, we considered the homologous temperature relation
\begin{equation}
D = D_{o} \exp ( g T_m /  T  )
\label{homo}
\end{equation}
commonly used to model the dependence of diffusion on $P$ and $T$\cite{sammis}.
Because of the similarity of behavior in diffusion of Mg and O, we
used the effective diffusion coefficient, $D_{eff} = 2 D_{Mg}
D_{O} / ( D_{Mg} + D_{O} )$ for $D$\cite{poirier}.
We tested this model using the
theoretical melting curve of Cohen and Weitz
\cite{cohen-weitz} obtained with the same VIB potential as used here,
and the extrapolated experimental melting curve of Zerr and Boehler
\cite{zerr-boe} which has a lower ${\rm d}P/{\rm d}T$.
Good global fits were found
using the theoretical melting curve, but the experimental melting
curve is not consistent with the present diffusion results.
As discussed in Ref. \onlinecite{cohen-weitz} the experimental
results may be influenced by Ar solubility in MgO melt at high
pressures. Theoretical estimates of the melting curve are generally
consistent with each other and with expected thermodynamic
parameters.

We also tested the use of zero pressure diffusion
results only in Eq.~\ref{homo} and found that extrapolations to high
pressure using the melting curve were reasonably reliable
although some accuracy was lost compared to the
results from direct high pressure simulations.  This gives justification
for use of melting curves in estimating high $P$ and $T$
diffusion in oxides. In addition, we found that $g$ was less than 14
while the average value for alkali halides is 24\cite{karato2} 
indicating that conclusions based
on a systematic value for this parameter may be invalid.

In summary, we found (1) excellent agreement with experimental results, (2)
that defects are formed from Schottky pairs as opposed to neutral divacancies,
and (3) the homologous temperature relation holds within our theoretical
framework. These results will help constrain rheological properties of the
deep Earth and provide constraints for pressure effects on kinetics in
oxides.

\begin{table}
\squeezetable
\caption{Free Energies of Formation and Migration and 
Cell Parameter\label{fe-dif}}
\begin{tabular}{ddd@{ $\pm$ }lc*{2}{l@{ $\pm$ }l}}
$P$&$T$&\multicolumn{2}{c}{$\Delta G_f$}&$a$&\multicolumn{4}{c}{$\Delta G_m$}\\
(GPa)&(K)&\multicolumn{2}{c}{($10^{-19}$ J)}&\multicolumn{1}{c}{(\AA)}&\multicolumn{4}{c}{($10^{-19}$J)}\\
  &  &\multicolumn{2}{c}{}&\multicolumn{1}{c}{}&\multicolumn{2}{c}{Mg}&\multicolumn{2}{c}{O}\\ \hline
0&1000&8.19&0.39&4.2495&3.36&0.10&3.75&0.11\\
0&2000&7.88&0.23&4.3073&2.72&0.12&3.16&0.12\\
20&2000&12.04&0.40&4.1312&3.75&0.10&4.09&0.10\\ 
20&3000&11.57&0.56&4.1687&3.60&0.10&4.06&0.12\\ 
80&2000&21.20&0.60&3.8610&5.16&0.10&5.84&0.17\\ 
80&3000&21.01&0.90&3.8803&4.99&0.11&5.86&0.07\\ 
80&4000&20.75&0.71&3.9031&4.55&0.10&5.61&0.16\\ 
140&2000&26.96&0.82&3.7071&6.39&0.10&6.91&0.10\\ 
140&3000&25.29&1.11&3.7210&5.99&0.15&6.21&0.21\\
140&5000&23.44&1.76&3.7509&4.80&0.30&5.69&0.41\\ 
\end{tabular}
\end{table}

\acknowledgments{We thank W.P. Reinhardt and V. Heine for helpful
discussions. This work was supported by NSF grant EAR94-18934. Computations
were performed on the Cray J916/12-1024 at the Geophysical Laboratory, CIW,
purchased with support from NSF grant EAR95-12627.}

%
%

%

\end{document}